\documentclass[dvips]{article}
\usepackage{txfonts}
\usepackage{graphicx}
\usepackage{natbib}
\usepackage[utf8]{inputenc}

\usepackage{icrctc07}

\def\intgr{{\it INTEGRAL}}

\title{A particle acceleration site in the Coma cluster?}
\authors{D. Eckert$^{1}$, N. Produit$^{1}$, A. Neronov$^{1}$ \& T. J.-L. Courvoisier$^{1}$}
\shortauthors{D. Eckert et al}
\afiliations{$^1$ISDC, 16, ch. d'Ecogia, Geneva Observatory, University of Geneva, CH-1290 Versoix, Switzerland}
\email{Dominique.Eckert@obs.unige.ch}

\abstract{We present the results of a deep (1.1 Ms) observation of the Coma cluster of galaxies in the 18-30 keV band with the ISGRI imager on board the \emph{INTEGRAL} satellite. We show that the source extension  in the North-East to South-West (SW) direction ($\sim 17'$)  significantly exceeds the size of the point spread function of ISGRI, and that the centroid of the image of the source in the 18-30~keV band is displaced in the SW direction compared to the centroid in the 1-10~keV band. To test the nature of the SW extension we fit the data assuming different models of source morphology. The best fit is achieved with a diffuse source of elliptical shape, although an acceptable fit can be achieved assuming an additional point source SW of the cluster core. In the case of an elliptical source, the direction  of extension of the source coincides with the direction toward the subcluster falling onto the Coma cluster. If the SW excess is due to the presence of a point source with a hard spectrum, we show that there is no obvious X-ray counterpart for this additional source, and that the closest X-ray source is the quasar EXO 1256+281, which is located $6.1'$ from the centroid of the excess. Finally, we show that the hard X-ray emission coincides with the 1.4 GHz radio emission, which suggests that the hard X-ray emission comes from the same population of electrons that is responsible for radio haloes through synchrotron emission.}

\begin{document}
\maketitle

\section{Introduction}
In the hierarchical scenario of structure formation, clusters of galaxies are the latest and biggest structures to form. Hence, we expect some of them to be still forming, and experiencing major merging events with smaller clusters. This is the case of the Coma cluster, that is currently merging with the NGC 4839 group.

In such events, the merging of the ICM of the two clusters creates shock fronts, in which theory predicts that an important population of particles would be accelerated to high energies \citep{sarazin}. This phenomenon should then produce a reheating of the gas, and create a higher temperature plasma that would radiate more strongly in hard X-rays. Alternatively, interaction of the population of mildly relativistic electrons that produce the halos of galaxy clusters via synchrotron radiation \citep{feretti} with the Cosmic Microwave Background would then produce hard X-ray emission through inverse Compton processes, and thus add a power-law tail to the spectrum in the hard X-ray domain. Detection of this hard X-ray excess would help to learn more about the cosmic ray population detected by radio observations. Furthermore, characterization of the morphology of the hard X-ray emission would bring a possible identification of acceleration sites.

Recent reports of detection of a hard X-ray excess by \emph{Beppo-SAX} \citep{fusco} and \emph{RXTE} \citep{rephaeli} in the Coma cluster seem to confirm the existence of a high energy tail of the spectrum of merging clusters, and thus prove the existence of particle acceleration sites in these clusters. However, these detections are rather weak and controversial \citep{rossetti}, and since the hard X-ray instruments on both \emph{Beppo-SAX} and \emph{RXTE} are non-imaging, contamination by very hard point sources inside the cluster could not be excluded (e.g. by the central galaxy NGC 4874, NGC 4889 or the QSO EXO 1256+281). Besides, it was not possible to have any information on the morphology of the hard X-ray emission.

\section{\intgr\ data analysis}

\begin{figure}
\resizebox{\hsize}{!}{\includegraphics{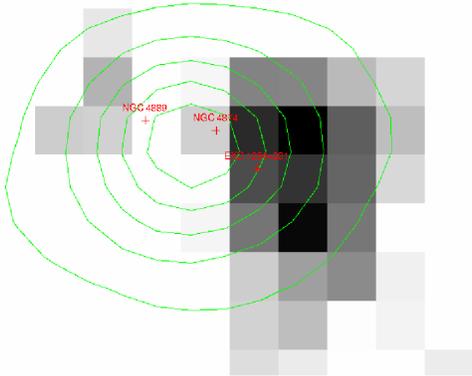}}
\caption{The residuals after the subtraction of the \emph{XMM-Newton} profile from the \emph{INTEGRAL} image (see text). The South-West excess in the residual image is apparent.}
\label{coma}
\end{figure}

After a careful selection of Science Windows (ScWs), we created a mosaic image of the cluster with the standard OSA 6.0 software in the 18-30 keV energy band (see \citet{eckert} for details). From this mosaic, we showed that the source extension  in the North-East to South-West (SW) direction ($\sim 17'$) significantly exceeds the size of the point spread function of ISGRI ($12'$ FWHM), and that the centroid of the image of the source in the 18-30~keV band is displaced in the SW direction compared to the centroid in the 1-10~keV band. Figure \ref{coma} shows  the residuals of the \emph{INTEGRAL} mosaic image after substraction of the smoothed \emph{XMM-Newton} image, renormalized in a way that the difference between \emph{INTEGRAL} and \emph{XMM-Newton} flux cancels at the maximum of the \emph{XMM-Newton} emission. One can clearly see that significant residuals are left in the South-West (SW) part of the \emph{INTEGRAL} source after the substraction. This indicates that the hard X-ray source detected by \emph{INTEGRAL} is more extended in the SW direction than the \emph{XMM-Newton} source.

\section{Possible interpretations for this excess}

To test the nature of the SW extension, we fitted the data assuming different models of source morphology. The best fit is achieved with a diffuse source of elliptical shape, with a semi major axis of $17'$. The direction of the major axis of the ellipse is inclined at the angle $\theta=61\pm 4^\circ$, which corresponds to the direction of the NGC 4839 group. Alternatively, an acceptable fit can be achieved assuming an additional point source SW of the cluster core. In this case, the additional point source is found at RA=$194.71\pm0.01$ and DEC=$27.87\pm0.01$. 

There is no obvious X-ray counterpart for this additional source: the closest X-ray source is the quasar EXO 1256+281, which is located $6.1'$ from the centroid of the excess (see Fig. \ref{coma}) for the position of this quasar). We also extracted the soft X-ray spectrum of this source and the flux of the south-west region in \emph{INTEGRAL} data, in order to check if the spectrum would be compatible with a highly absorbed Seyfert II galaxy. We find that this is very unlikely: indeed, the fit requires a very high absorption ($n_H>4\times 10^{24}\mbox{ cm}^{-2}$) and a steep spectral index $\Gamma\geq 3.0$ for the absorbed component, which would be very unusual for a Seyfert II-type Active Galactic Nucleus. We cannot exclude the possibility that this excess is due to an unknown X-ray point source, but because of the spectral properties required for this object, this hypothesis is very unlikely. As a conclusion, we claim that contribution of a very hard point source embedded in the cluster to the observed flux is highly unlikely. 

If we interpret the SW \intgr\ excess as diffuse emission, which gives the best representation of the data, we see that the source is extended towards the sub-cluster around NGC 4839, which gives an indication that the emission might be related to the currently on-going merger between the main cluster and this sub-cluster. We have investigated two possibilities for this additional component: a thermal Bremsstrahlung emission from a hotter region, and a non-thermal Inverse-Compton (IC) component from the $E\sim$GeV electrons that produce radio halos.

In the case of a hotter region, we find that this region coincides with a very hot region ($kT\geq10$ keV, \citet{neumann}). A joint fit of the of 1-10~keV spectrum extracted from a circle of a radius of 6' centered at the position of the SW excess and the 18-50~keV spectrum extracted from the SW region from the \intgr\ mosaic image give an upper limit to this temperature of $kT\leq14$ keV. This result is comparable with the temperature found in the merging region of the distant cluster Cl J0152.7-1357 \citep{maughan}, which shows that such a high temperature is possible and might indeed be the signature of a merger. We can thus associate the hard X-ray excess in this region with emission from a very hot region of the cluster ($10\mbox{ keV}\leq kT\leq14$ keV).

\begin{figure}
\resizebox{\hsize}{!}{\includegraphics{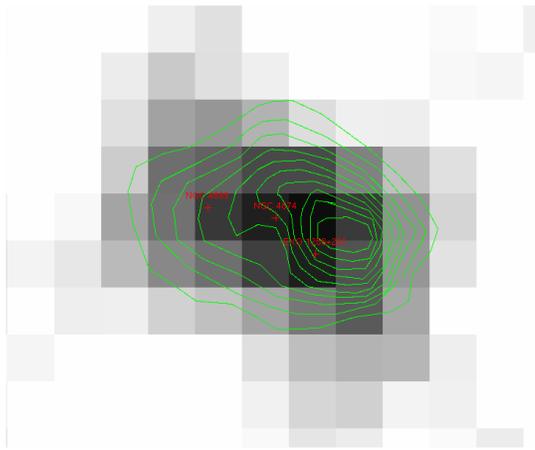}}
\caption{The \intgr\ significance image with 1.4 GHz radio contours overlayed.}
\label{radio}
\end{figure}

If the emission from the SW region comes from non-thermal IC scattering of midly relativistic electrons, which is the most popular idea to explain the high-energy tail detected by \emph{Beppo-SAX}, then the morphology of the hard X-ray emission must be correlated to the morphology of the radio halo of the cluster. Figure \ref{radio} shows the \intgr\ mosaic image in the 18-30 keV band with 1.4 GHz radio contours overlayed from \citet{deiss}. The radio contours are substracted for the radio galaxies. We can see that the radio emission is also displaced compared to the 1-10 keV thermal emission, and that the maximum of the radio emission coincides with the SW excess. This strongly suggests that the hard X-ray emission from this region is of non-thermal origin, namely of IC scattering from the same population of electrons as the radio emission. This strengthens the controversial result of \citet{fusco}.

\section{Conclusion}

Thanks to the imaging capabilities of ISGRI, we were able for the first time to resolve spatially the Coma cluster in the hard X-ray domain. We showed that the hard X-ray emission is displaced compared to the purely thermal emission in soft X-rays. Investigating this displacement, we found that the contribution of point sources to the observed spectrum is likely negligible. Our analysis shows that a hotter plasma can explain this excess. However, the correlation between the radio and hard X-ray morphology strongly suggests that the SW excess is due to IC scattering of mildly relativistic electrons.

Together with the spectral results obtained by \emph{Beppo-SAX} \citep{fusco2}, it is now becoming clear that the presence of an additional X-ray spectral component is required by the data. Moreover, the strong correlation between radio and hard X-ray morphology clarifies the nature of this excess, and seems to confirm the existence of a particle acceleration site in the Coma cluster.

\bibliographystyle{aa}
\bibliography{eckert_icrc}

\begin{thebibliography}{10}
\expandafter\ifx\csname natexlab\endcsname\relax\def\natexlab#1{#1}\fi

\bibitem[{{Deiss} {et~al.}(1997){Deiss}, {Reich}, {Lesch}, \&
  {Wielebinski}}]{deiss}
{Deiss}, B.~M., {Reich}, W., {Lesch}, H., \& {Wielebinski}, R. 1997, \aap, 321,
  55

\bibitem[{{Eckert} {et~al.}(2007){Eckert}, {Neronov}, {Courvoisier}, \&
  {Produit}}]{eckert}
{Eckert}, D., {Neronov}, A., {Courvoisier}, T.~J.-L., \& {Produit}, N. 2007,
  \aap, 470, 835

\bibitem[{{Feretti} \& {Giovannini}(2007)}]{feretti}
{Feretti}, L. \& {Giovannini}, G. 2007, ArXiv Astrophysics e-prints

\bibitem[{{Fusco-Femiano} {et~al.}(2007){Fusco-Femiano}, {Landi}, \&
  {Orlandini}}]{fusco2}
{Fusco-Femiano}, R., {Landi}, R., \& {Orlandini}, M. 2007, \apjl, 654, L9

\bibitem[{{Fusco-Femiano} {et~al.}(2004){Fusco-Femiano}, {Orlandini},
  {Brunetti}, {Feretti}, {Giovannini}, {Grandi}, \& {Setti}}]{fusco}
{Fusco-Femiano}, R., {Orlandini}, M., {Brunetti}, G., {et~al.} 2004, \apjl,
  602, L73

\bibitem[{{Maughan} {et~al.}(2003){Maughan}, {Jones}, {Ebeling}, {Perlman},
  {Rosati}, {Frye}, \& {Mullis}}]{maughan}
{Maughan}, B.~J., {Jones}, L.~R., {Ebeling}, H., {et~al.} 2003, \apj, 587, 589

\bibitem[{{Neumann} {et~al.}(2003){Neumann}, {Lumb}, {Pratt}, \&
  {Briel}}]{neumann}
{Neumann}, D.~M., {Lumb}, D.~H., {Pratt}, G.~W., \& {Briel}, U.~G. 2003, \aap,
  400, 811

\bibitem[{{Rephaeli} \& {Gruber}(2002)}]{rephaeli}
{Rephaeli}, Y. \& {Gruber}, D. 2002, \apj, 579, 587

\bibitem[{{Rossetti} \& {Molendi}(2004)}]{rossetti}
{Rossetti}, M. \& {Molendi}, S. 2004, \aap, 414, L41

\bibitem[{{Sarazin}(1999)}]{sarazin}
{Sarazin}, C.~L. 1999, ArXiv Astrophysics e-prints

\end{thebibliography}
\end{document}